\documentclass[11pt]{article}

\usepackage[english]{babel}
\usepackage{amsmath,amsthm}
\usepackage{amsfonts}
\usepackage{bm}
\usepackage{float}
\usepackage{caption}
\usepackage{a4wide}
\usepackage{graphicx}
\usepackage[list=true]{subcaption}
\usepackage{algorithm}
\usepackage[titletoc]{appendix}
\usepackage{xcolor}
\usepackage{cite}

\usepackage{fancyhdr}  
\usepackage{lastpage}  
\usepackage{array}
\usepackage{hyperref}
\hypersetup{
    colorlinks=true,
    linkcolor=blue,
    filecolor=magenta,
    urlcolor=blue,
}
\DeclareMathOperator*{\argmin}{argmin}

\DeclareMathOperator{\diver}{div}
\newcommand{\tran}{^{\mathsf{T}}}

\newcommand{\diff}{\mathrm{d}}
\newcommand{\Diff}{\mathrm{D}}
\newcommand{\Sc}{\mathcal{S}}
\newcommand{\Tc}{\mathcal{T}}
\newcommand{\Xc}{\mathcal{X}}
\newcommand{\Zc}{\mathcal{Z}}
\newcommand{\sray}{\bm{\hat{s}}}
\newcommand{\tray}{\bm{\hat{t}}}
\newcommand{\nvec}{\bm{\hat{n}}}
\newcommand{\bom}{\bm{\omega}}

\title{Design of a freeform two-reflector system to collimate and shape a point source distribution}
\author{A.H. van Roosmalen,\textsuperscript{1,*} \and M.J.H. Anthonissen,\textsuperscript{1} \and W.L. IJzerman,\textsuperscript{1,2} \and J.H.M. ten Thije Boonkkamp\textsuperscript{1}}
\date{}

\begin{document}
\maketitle

\noindent \textsuperscript{1}CASA, Department of Mathematics and Computer Science, Eindhoven University of \\Technology, PO Box 513, 5600 MB Eindhoven, The Netherlands\\
\textsuperscript{2}Signify Research, High Tech Campus 7, 5656 AE Eindhoven, The Netherlands\\
\textsuperscript{*}Corresponding author: \href{mailto:a.h.v.roosmalen@tue.nl}{a.h.v.roosmalen@tue.nl}

\begin{abstract}
  \noindent In this paper we propose a method to compute a freeform reflector system for collimating and shaping a beam from a point source. We construct these reflectors such that the radiant intensity of the source is converted into a desired target. An important generalization in our approach compared to previous research is that the output beam can be in an arbitrary direction. The design problem is approached by using a generalized Monge-Amp\`ere equation. This equation is solved using a least-squares algorithm for non-quadratic cost functions. This algorithm calculates the optical map, from which we can then compute the surfaces. We test our algorithm on two cases. First we consider a uniform source and target distribution. Next, we use the model of a laser diode light source and a ring-shaped target distribution.
\end{abstract}

\section{Introduction}
Beam shaping is an important research topic within illumination optics. Especially the shaping of a parallel beam into another parallel beam with a different distribution is well researched \cite{oli07,yad18,dic18}. This is often linked to the shaping of laser beams. A common source for a laser is a laser diode, which can be modelled as a point source \cite{li92,sun15}. The diverging beam from such a diode is often collimated first with a lens, before other manipulations such as beam shaping are applied \cite{sun15}. An optical system that directly shapes and collimates the output from a point source is able to skip the first collimation step. This can reduce the total number of necessary optical surfaces and increase the efficacy by avoiding Fresnel reflections. An example of a useful light distribution in a collimated beam that we will discuss is the ring shape, meaning that the projection of this beam on a plane perpendicular to it gives a ring-shaped illumination pattern. A possible benefit of such a ring-shaped target is that a subsequent focussing can be done more accurately \cite{fuc15}. A ring-shaped illumination pattern also has a use in welding \cite{las15}. \\
\indent In most research regarding collimated beams, it is assumed that the outgoing beam is in the same direction as the incoming beam. For the case of a point source, however, there is not one single direction of light emission, but often there exists a symmetry axis that plays the role of beam direction. In this paper we will drop that assumption and allow the outgoing beam to be in any arbitrary direction. This gives us the possibility to create so-called folded optics. As a result we can design more compact optical systems. As mentioned, we will look for an optical system to collimate a beam from a point source. For this we need two optical surfaces, one to shape the light to the desired target distribution and one to collimate the beam. We choose to work with two reflector surfaces as freeform optical surfaces. \\
\indent Optical design for illumination can roughly be divided into two categories: forward and inverse methods. The former deals with the calculation of the output of an optical system. The result of such a forward method can then be used to iteratively refine the design, which is a slow process \cite{fil21}. In this paper we will focus on inverse methods. With such methods, the goal is to compute the shape of the optical system given source and target illumination patterns. \\
\indent Although the example of a laser gives an idea of the possibilities, we will not restrict ourselves to this. In our derivations we will use the approximation of geometrical optics. This way we can view the calculation of the surfaces as an optimal transport problem, i.e., to find an optical map that `transports' the light from the source to the target. This gives us a Monge-Amp\`ere type equation with transport boundary condition \cite{yad18}. \\
\indent Several methods have been developed for solving problems similar to the point source and parallel outgoing beam. A more thorough overview of inverse methods can be found in \cite{rom19}. Some of them use numerical methods such as finite differences and Newton's method to directly solve the Monge-Amp\`ere equation \cite{wu13,bos17}. Oliker et al. proposed the supporting quadric method \cite{oli07,oli15}. Alternatively, the optimal mass transport problem is reduced to a linear assignment problem by Doskolovich et al. \cite{dosko19}. Another approach by Feng et al. uses ray mapping to calculate the shapes and positions of the surfaces \cite{feng15}. A ray mapping method has been used to construct optical systems with two freeform surfaces for arbitrary input and output wavefronts \cite{wei19}. \\
\indent To the best of our knowledge, none of the above approaches have been used to specifically design an optical system for collimating and shaping a diverging beam from a point source. In this paper we will modify a least-squares method to solve the Monge-Amp\`ere type equation. Versions of this algorithm have been used before for multiple optical design challenges, including parallel to far-field \cite{prins15}, parallel to parallel \cite{yad19} and point source to far-field \cite{rom20}. \\
\indent In Section 2 we present the mathematical model linking the shapes of the surfaces to the source and target distributions. The equations in this model are used in the algorithm mentioned before. We will give a short summary of this algorithm in Section 3. For some parts of the algorithm we refer to other papers and we briefly discuss the most important parts. In Section 4 we test our algorithm on two test cases: First a uniform point source and a uniform target, second a laser diode source and a ring-shaped target. The conclusion of our findings is given in Section 5.

\section{Formulation of the mathematical model}\label{sec:problem formulation}
\begin{figure}[ht!]
  \centering
  \def\svgwidth{\linewidth}
  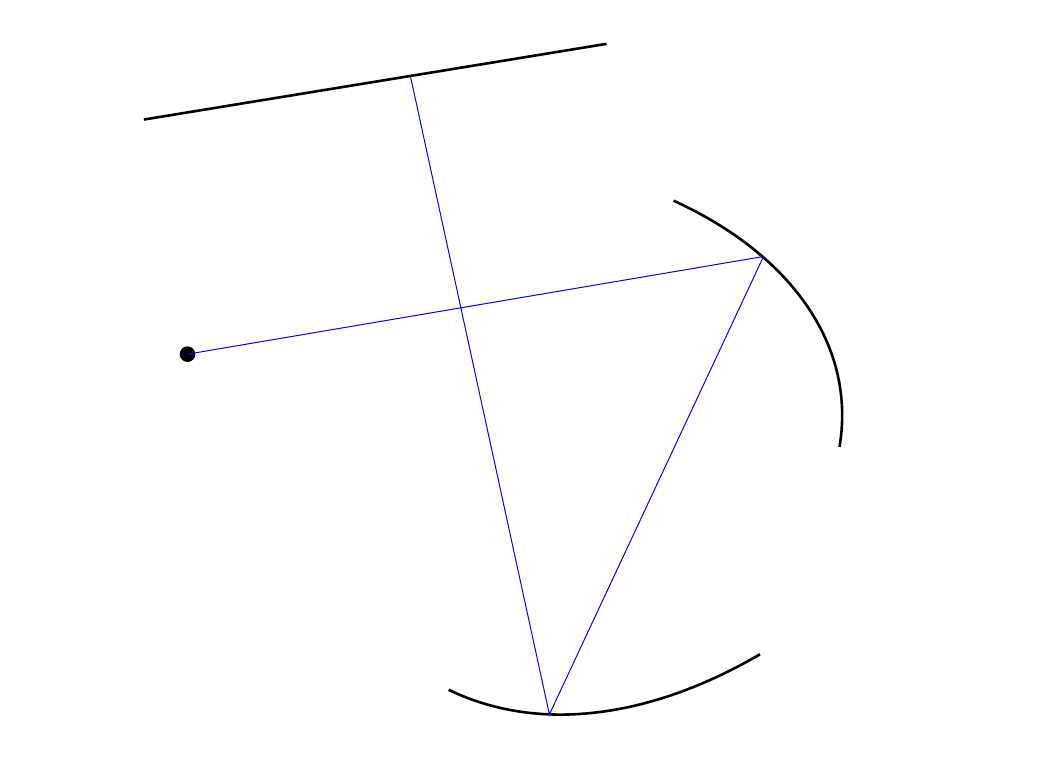
  \caption{A sketch of the optical system in the $x,z$-plane.}\label{fig:overview}
\end{figure}

In this section we formulate the mathematical model of a reflective optical system creating a collimated beam from a point source. Given a source light distribution we want to design an optical system of two reflectors. The first reflector will be used to shape the intensity profile and the second one will collimate the beam. The collimated output beam should give a light distribution on a target plane at a distance $l$ from the source, where the distribution is a given function of the position coordinates in the plane. A two-dimensional illustration of the system is given in Fig. \ref{fig:overview}.
\subsection{Derivation of the cost function}
We choose the point source to be located at the origin of a coordinate system given by the standard basis $\bm{\hat{e}}_1$, $\bm{\hat{e}}_2$, $\bm{\hat{e}}_3$. The orientation for this coordinate system is arbitrary. Often, the $\bm{\hat{e}}_3$-axis is chosen to coincide with the symmetry axis of the bundle of light emitted from the source. However, such a symmetry does not necessarily exist and will not be assumed for our derivations. The only restriction we have on the coordinate system is that there can be no light emitted in the negative $\bm{\hat{e}}_3$-direction. The reason for this will become apparent later. This source emits rays with unit direction vectors $\sray=(s_1,s_2,s_3)\tran$. Two reflectors, $\mathcal{R}_1$ and $\mathcal{R}_2$, are used to collimate and shape the light emitted by the source into an output beam with the desired intensity distribution. These output rays should propagate as a parallel beam that is parallel to a given direction vector $\bm{\hat{a}}_3$ perpendicular to the target plane. The direction of $\bm{\hat{a}}_3$ with respect to $\bm{\hat{e}}_1,\dots,\bm{\hat{e}}_3$ is given by a rotation of $\bm{\hat{e}}_3$ with a polar angle $\varphi$ around the $\bm{\hat{e}}_2$-axis and subsequently with an azimuthal angle $\theta$ around the $\bm{\hat{e}}_3$-axis. The matrix associated with this composite rotation is given by
\begin{equation}
  \bm{A}=\begin{pmatrix}
                    \cos\varphi\cos\theta & -\sin\theta & \sin\varphi\cos\theta \\
                    \cos\varphi\sin\theta & \cos\theta & \sin\varphi\sin\theta \\
                    -\sin\varphi & 0 & \cos\varphi
                  \end{pmatrix},
\end{equation}
and we define $\bm{\hat{a}}_i=\bm{A}\bm{\hat{e}}_i$, $i=1,2,3$. These $\bm{\hat{a}}_i$-vectors are the rotations of an orthogonal basis, so they are again an orthogonal basis and $\bm{\hat{a}}_i$ is equal to the $i$th column of $\bm{A}$. The target plane is perpendicular to $\bm{\hat{a}}_3$ at a distance $l$ from the point source along $\bm{\hat{a}}_3$. Any point on the target plane is then given by $y_1\bm{\hat{a}}_1+y_2\bm{\hat{a}}_2+l\bm{\hat{a}}_3$. The position in the target plane is denoted by $\bm{y}=(y_1,y_2)\tran$. The first reflector, $\mathcal{R}_1$, is defined by the radial distance from the source, $u=u(\sray)$. The second reflector, $\mathcal{R}_2$, is given by the perpendicular distance from the target plane, $w=w(\bm{y})$. The reflectors are then mathematically described by
\begin{equation}\label{eq:reflectors}
  \mathcal{R}_1 : \bm{r}_1(\sray) = u(\sray)\sray, \qquad \mathcal{R}_2 : \bm{r}_2(\bm{y}) = y_1\bm{\hat{a}}_1+y_2\bm{\hat{a}}_2+(l-w(\bm{y}))\bm{\hat{a}}_3 = \bm{A}\bm{r}'_2(\bm{y}),
\end{equation}
where $\bm{r}'_2(\bm{y})=\big(y_1,y_2,l-w(\bm{y})\big)\tran$. With $P$ and $Q$ we denote the points where a ray hits the first and second reflector, respectively, see Fig. \ref{fig:overview}. The distance between those two points is denoted by $d$. In the framework of optimal mass transport we want to derive a relation between $\sray$ and $\bm{y}$ of the form
\begin{equation}\label{eq:costform}
  u_1(\sray)+u_2(\bm{y})=c(\sray,\bm{y}),
\end{equation}
where $u_1$ and $u_2$ are related to the shape and location of the surfaces and $c$ is the so-called cost function \cite{yad18}. For any ray, we introduce the following notation. The 2-vectors $\bm{q}_s=\bm{0}$ and $\bm{q}_t=\bm{y}$ denote the position of the ray at the source and target, respectively. These are projections of 3-dimensional position vectors onto the plane $z=0$ (source plane) and the target plane, respectively. Similarly, $\bm{p}_s$ and $\bm{p}_t=\bm{0}$ are the projections of the direction vectors on the source and target planes. In terms of Hamiltonian characteristics, the optical path length (OPL), denoted by $L$, is equal to the point characteristic $V$\cite{lun64}. We can write
\begin{equation}\label{eq:opl}
  V(\bm{q}_s,\bm{q}_t) = u(\sray)+d+w(\bm{y}).
\end{equation}
Because the variables $\sray$ and $\bm{y}$ denote a direction at the source and a position on the target, we work with the second mixed characteristic $W^*=W^*(\bm{p}_s,\bm{q}_t)$, given by
\begin{equation}
  W^*(\bm{p}_s,\bm{q}_t) = V(\bm{q}_s,\bm{q}_t) + \bm{q}_s\bm{\cdot}\bm{p}_s.
\end{equation}
However, since $\bm{q}_s=\bm{0}$ we can dismiss the second term and the mixed characteristic is equal to the optical path length. The following relations can be derived for the mixed characteristic \cite{lun64}
\begin{equation}
  \frac{\partial W^*}{\partial\bm{p}_s} = \bm{q}_s = \bm{0},\qquad\frac{\partial W^*}{\partial\bm{q}_t} = \bm{p}_t = \bm{0}.
\end{equation}
This proves that $W^*$ is independent of the direction from the source and the position on the target plane. As a result, the optical path length $L=V=W^*$ is a constant. \\
\indent We will eliminate $d$ from Eq. (\ref{eq:opl}) by using the fact that $d$ is the distance between $P$ and $Q$. We denote by $\bm{r}_P=\bm{r}_1(\sray)$ and $\bm{r}_Q=\bm{r}_2(\bm{y})$ the position vectors of $P$ and $Q$, respectively. Furthermore, we use $\bm{r}_Q'=\bm{r}_2'(\bm{y})$, so we can write
\begin{equation}
\begin{aligned}
  d^2 &= |\bm{r}_Q-\bm{r}_P|^2 \\
  &= |\bm{A}\bm{r}'_Q-\bm{r}_P|^2 \\
  &= |\bm{A}\bm{r}'_Q|^2 + |\bm{r}_P|^2 - 2\left(\bm{A}\bm{r}'_Q\right)\bm{\cdot}\bm{r}_P.
\end{aligned}
\end{equation}
Because $\bm{A}$ is a rotation matrix, we have $|\bm{A}\bm{r}'_Q|=|\bm{r}'_Q|$. We also have $\bm{r}_P=u\sray$ and $\bm{A} = (\bm{\hat{a}}_1,\bm{\hat{a}}_2,\bm{\hat{a}}_3)$. This can be used to write
\begin{equation}\label{eq:dist1}
  \begin{aligned}
  d^2 &= |\bm{r}'_Q|^2 + u^2 - 2u\left(\bm{A}\bm{r}'_Q\right)\bm{\cdot}\sray \\
  &= u^2 + |\bm{y}|^2 +(l-w)^2 -2u\big(y_1\bm{\hat{a}}_1+y_2\bm{\hat{a}}_2+(l-w)\bm{\hat{a}}_3\big)\bm{\cdot}\sray \\
  &= u^2 + |\bm{y}|^2 +l^2-2lw+w^2 -2u\sray\bm{\cdot}(\bm{A}_2\bm{y})-2ul\sray\bm{\cdot}\bm{\hat{a}}_3 + 2uw\sray\bm{\cdot}\bm{\hat{a}}_3,
  \end{aligned}
\end{equation}
where $\bm{A}_2 = (\bm{\hat{a}}_1,\bm{\hat{a}}_2)\in\mathbb{R}^{3\times2}$ and we omit the dependence of $u$ and $w$ on $\sray$ and $\bm{y}$ for now. We combine Eq. (\ref{eq:opl}) and Eq. (\ref{eq:dist1}) to obtain
\begin{equation}\label{eq:interm1}
  |\bm{y}|^2+l^2-2lw-2u\sray\bm{\cdot}(\bm{A}_2\bm{y})-2ul\sray\bm{\cdot}\bm{\hat{a}}_3+ 2uw\sray\bm{\cdot}\bm{\hat{a}}_3-L^2+2Lu+2Lw-2uw = 0.
\end{equation}
We want to separate $u$ and $w$ to get an equation of the form of Eq. (\ref{eq:costform}). For that, we first divide Eq. (\ref{eq:interm1}) by $u$. It is reasonable to assume $u>0$, since otherwise the reflector coincides with the source at some point. Introducing $\tilde{u}=1/u$, Eq. (\ref{eq:interm1}) reads
\begin{equation}
  |\bm{y}|^2\tilde{u}+l^2\tilde{u}-2l\tilde{u}w-2\sray\bm{\cdot}(\bm{A}_2\bm{y})-2l\sray\bm{\cdot}\bm{\hat{a}}_3+ 2w\sray\bm{\cdot}\bm{\hat{a}}_3-L^2\tilde{u}+2L+2L\tilde{u}w-2w = 0.
\end{equation}
The parameter $\beta=L-l$ is introduced, which is called the reduced optical path length. This is used to rewrite the equation above as
\begin{equation}
  \tilde{u}w+\tilde{u}\frac{1}{2}\left(\frac{1}{\beta}|\bm{y}|^2-(L+l)\right)-w\frac{1}{\beta}\big(1-\sray\bm{\cdot}\bm{\hat{a}}_3\big) =\frac{\sray\bm{\cdot}(\bm{A}_2\bm{y})+l\sray\bm{\cdot}\bm{\hat{a}}_3-L}{\beta}.
\end{equation}
To obtain an equation of the desired form, we will factorize the equation above and then apply logarithms to both sides of the equation. The factorization leads to
\begin{subequations}\label{eq:kappas}
  \begin{align}
  &\kappa_1(\sray)\kappa_2(\bm{y}) = T(\sray,\bm{y}), \label{eq:kappas1}\\
  &\kappa_1(\sray) = \tilde{u}-\frac{1}{\beta}\left(1-\sray\bm{\cdot}\bm{\hat{a}}_3\right),\qquad \kappa_2(\bm{y})=w+\frac{1}{2}\left(\frac{|\bm{y}|^2}{\beta}-(L+l)\right), \\
  &T(\sray,\bm{y}) = \frac{1}{\beta}\sray\bm{\cdot}\bm{A}_2\bm{y}-\frac{1-\sray\bm{\cdot}\bm{\hat{a}}_3}{2\beta^2}|\bm{y}|^2-\frac{1}{2}\left(\sray\bm{\cdot}\bm{\hat{a}}_3+1\right).
  \end{align}
\end{subequations}
As mentioned, we would like to take the logarithm of both sides of the equation to get to the form of Eq. (\ref{eq:costform}). However, we need to make sure that both sides are positive. We use $L=u+d+w$ and we can write
\begin{equation}\label{eq:l expanded}
  l = u\sray\bm{\cdot}\bm{\hat{a}}_3+d\tray\bm{\cdot}\bm{\hat{a}}_3+w,
\end{equation}
where $\tray$ is the (unit) direction vector after the first reflection. Combining both relations, we obtain $\beta = (1-\sray\bm{\cdot}\bm{\hat{a}}_3)u+(1-\tray\bm{\cdot}\bm{\hat{a}}_3)d$. Since $\tray$ is a unit direction vector, we know that $\beta\geq(1-\sray\bm{\cdot}\bm{\hat{a}}_3)u\geq0$, with the first equality only if $\tray\bm{\cdot}\bm{\hat{a}}_3=1$. In that case $\tray$ is parallel to $\bm{\hat{a}}_3$, so there would be no second reflection. We can disregard this case and therefore write
\begin{equation}
  \kappa_1 = \frac{1}{u}-\frac{1-\sray\bm{\cdot}\bm{\hat{a}}_3}{\beta} > \frac{1}{u}-\frac{1-\sray\bm{\cdot}\bm{\hat{a}}_3}{(1-\sray\bm{\cdot}\bm{\hat{a}}_3)u} = 0.
\end{equation}
The vector $\bm{y}$ is the displacement along the plane spanned by $\bm{\hat{a}}_1$ and $\bm{\hat{a}}_2$. We have $y_1\bm{\hat{a}}_1+y_2\bm{\hat{a}}_2=\bm{A}_2\bm{y}$. This displacement is determined completely by the first two ray segments; from the source to the first reflector, and from the first to the second reflector. Using this, we write
\begin{equation}\
  y_1 = u\sray\bm{\cdot}\bm{\hat{a}}_1+d\tray\bm{\cdot}\bm{\hat{a}}_1,\qquad y_2 = u\sray\bm{\cdot}\bm{\hat{a}}_2+d\tray\bm{\cdot}\bm{\hat{a}}_2.
\end{equation}
Substituting this into the expression for $\kappa_2$ and using Eq. (\ref{eq:opl}) and Eq. (\ref{eq:l expanded}) gives us
\begin{equation}
\begin{aligned}
  2\beta\kappa_2 &= 2\beta w+\left(u\sray\bm{\cdot}\bm{\hat{a}}_1+d\tray\bm{\cdot}\bm{\hat{a}}_1\right)^2+\left(u\sray\bm{\cdot}\bm{\hat{a}}_2+d\tray\bm{\cdot}\bm{\hat{a}}_2\right)^2 \\ &\qquad-\beta\left((1+\sray\bm{\cdot}\bm{\hat{a}}_3)u+(1+\tray\bm{\cdot}\bm{\hat{a}}_3)d+2w\right).
\end{aligned}
\end{equation}
We now use again that $\beta = (1-\sray\bm{\cdot}\bm{\hat{a}}_3)u+(1-\tray\bm{\cdot}\bm{\hat{a}}_3)d$ to obtain
\begin{equation}
\begin{aligned}
  2\beta\kappa_2 &= \left(u\sray\bm{\cdot}\bm{\hat{a}}_1+d\tray\bm{\cdot}\bm{\hat{a}}_1\right)^2+\left(u\sray\bm{\cdot}\bm{\hat{a}}_2+d\tray\bm{\cdot}\bm{\hat{a}}_2\right)^2 \\
  &\qquad-\left((1-\sray\bm{\cdot}\bm{\hat{a}}_3)u+(1-\tray\bm{\cdot}\bm{\hat{a}}_3)d\right)\left((1+\sray\bm{\cdot}\bm{\hat{a}}_3)u+(1+\tray\bm{\cdot}\bm{\hat{a}}_3)d\right) \\
  &= u^2\left(-1+\sum_{i=1}^3(\sray\bm{\cdot}\bm{\hat{a}}_i)^2\right)+d^2\left(-1+\sum_{i=1}^3(\tray\bm{\cdot}\bm{\hat{a}}_i)^2\right) -ud\left(2-2\sum_{i=1}^3(\sray\bm{\cdot}\bm{\hat{a}}_i)(\tray\bm{\cdot}\bm{\hat{a}}_i)\right).
\end{aligned}
\end{equation}
Note that we can combine the $\sray\bm{\cdot}\bm{\hat{a}}_i$ terms into a vector written as
\begin{equation}
  \begin{pmatrix}
    \sray\bm{\cdot}\bm{\hat{a}}_1 \\
    \sray\bm{\cdot}\bm{\hat{a}}_2 \\
    \sray\bm{\cdot}\bm{\hat{a}}_3
  \end{pmatrix} = \bm{A}\tran\sray.
\end{equation}
This vector can be used to eliminate the $u^2$ term, since
\begin{equation}
  \sum_{i=1}^3(\sray\bm{\cdot}\bm{\hat{a}}_i)^2 = |\bm{A}\tran\sray|^2 = |\sray|^2 = 1,
\end{equation}
as $\bm{A}$ is a rotation matrix and therefore $\bm{A}\bm{A}\tran=\bm{I}$. Similarly, swapping $\tray$ for $\sray$ gives us $\sum_{i=1}^3(\tray\bm{\cdot}\bm{\hat{a}}_i)^2=1$. We use this to rewrite
\begin{equation}
\begin{aligned}
  2\beta\kappa_2 &= -2ud\left(1-\sum_{i=1}^3(\sray\bm{\cdot}\bm{\hat{a}}_i)(\tray\bm{\cdot}\bm{\hat{a}}_i)\right) \\
  &= -2ud\left(1-(\bm{A}\tran\sray)\bm{\cdot}(\bm{A}\tran\tray)\right) \\
  &= -2ud\left(1-\sray\bm{\cdot}\tray\right) \\
  &\leq 0,
\end{aligned}
\end{equation}
with equality if and only if $\tray=\sray$. This will not occur, since there would be no reflection at the first surface in that case. We deduced that $\kappa_1>0$ and $\kappa_2<0$. Consequently, $T$ must be negative as well. Both sides of Eq. (\ref{eq:kappas1}) are therefore multiplied by $-1$ to obtain
\begin{equation}\label{eq:kappa pos}
\kappa_1(\sray)(-\kappa_2(\bm{y}))=-T(\sray,\bm{y}).
\end{equation}
Before we apply the logarithm to this equation, we scale all the lengths by a factor $\beta$. Note that $\sray$ is already dimensionless. We introduce the variable $\bm{z}$ such that $\bm{y}=\beta\bm{z}$. We substitute this into the function $-T(\sray,\bm{y})=\hat{T}(\sray,\bm{z})$ and obtain
\begin{equation}
  \hat{T}(\sray,\bm{z}) = -\sray\bm{\cdot}\bm{A}_2\bm{z}+\frac{1}{2}\left(1-\sray\bm{\cdot}\bm{\hat{a}}_3\right)|\bm{z}|^2+\frac{1}{2}\left(\sray\bm{\cdot}\bm{\hat{a}}_3+1\right).
\end{equation}
Next, we scale all the other lengths, viz. $w(\bm{y})=\beta\hat{w}(\bm{z})$, $L=\beta\hat{L}$ and $l=\beta\hat{l}$. Furthermore, $\tilde{u}$ has the dimension of length inverse, so we scale this by $1/\beta$ to $\tilde{u}(\sray)=(1/\beta)\hat{u}(\bm{x})$. Substituting this in the expressions for $\kappa_1$ and $\kappa_2$ gives
\begin{subequations}
\begin{align}
  \kappa_1(\sray) &= \frac{1}{\beta}\big(\hat{u}(\bm{x})+\sray\bm{\cdot}\bm{\hat{a}}_3-1\big), \\
  \kappa_2(\bm{y}) &= \beta\left(\hat{w}(\bm{z})+\frac{1}{2}\left(|\bm{z}|^2-(\hat{L}+\hat{l})\right)\right).
\end{align}
\end{subequations}
So we can now define the new functions $\hat{\kappa}_1$ and $\hat{\kappa}_2$ by
\begin{subequations}\label{eq:kappa defs}
\begin{align}
  \hat{\kappa}_1(\sray) &= \hat{u}(\sray)+\sray\bm{\cdot}\bm{\hat{a}}_3-1, \\
  \hat{\kappa}_2(\bm{z}) &= \hat{w}(\bm{z})+\frac{1}{2}\left(|\bm{z}|^2-(\hat{L}+\hat{l})\right),
\end{align}
\end{subequations}
and $\kappa_1\kappa_2=\hat{\kappa}_1\hat{\kappa}_2$. Now Eq. (\ref{eq:kappa pos}) becomes
\begin{equation}\label{eq:kappa scaled}
  \hat{\kappa}_1(\sray)\left(-\hat{\kappa}_2(\bm{z})\right) = \hat{T}(\sray,\bm{z}).
\end{equation}
For the algorithm that will be introduced in Section \ref{sec:algorithm}, we want to change the source coordinates into two independent variables instead of three variables on the unit sphere. For that we choose the stereographic projection from the south pole $(0,0,-1)$ onto the equator plane $z=0$, written as $\bm{x}$. These coordinates are given by
\begin{equation}
  \bm{x}=\begin{pmatrix}x_1 \\ x_2\end{pmatrix}=\frac{1}{1+s_3}\begin{pmatrix}s_1 \\ s_2\end{pmatrix}, \qquad
  \sray=\begin{pmatrix}s_1 \\ s_2 \\ s_3\end{pmatrix}=\frac{1}{|\bm{x}|^2+1}\begin{pmatrix}2x_1 \\ 2x_2 \\ 1-|\bm{x}|^2\end{pmatrix}.
\end{equation}
Furthermore, we introduce
\begin{subequations}\label{eq:shapes}
  \begin{align}
  u_1(\bm{x}) &= \log\left(\hat{\kappa}_1\big(\sray(\bm{x})\big)\right),\\
  u_2(\bm{z}) &= \log\left(-\hat{\kappa}_2(\bm{z})\right).
  \end{align}
\end{subequations}
When we take the logarithm of both sides of Eq. (\ref{eq:kappa scaled}), taking into account the functions we just introduced, we get the desired form of Eq. (\ref{eq:costform}), given by
\begin{equation}\label{eq:main1}
  u_1(\bm{x})+u_2(\bm{z}) = c(\bm{x},\bm{z}),
\end{equation}
where the function $c$ is called the cost function and is defined as
\begin{equation}
\begin{aligned}
  c(\bm{x},\bm{z}) &= \log\left(\hat{T}\left(\sray(\bm{x}),\bm{z}\right)\right) \\
   &=\log\left[-\sray(\bm{x})\bm{\cdot}\bm{A}_2\bm{z}+\frac{1}{2}\Big(1-\sray(\bm{x})\bm{\cdot}\bm{\hat{a}}_3\Big)|\bm{z}|^2+\frac{1}{2}\Big(\sray(\bm{x})\bm{\cdot}\bm{\hat{a}}_3+1\Big)\right].
\end{aligned}
\end{equation}

\subsection{Energy conservation}\label{sec:energy}
We have deduced an equation that implicitly links $\bm{x}$ and $\bm{z}$. We assume that there exists an explicit relation given by $\bm{z}=\bm{m}(\bm{x})$. A constraint for this function is that the energy of the source should be conserved through the optical system. By $\Sc\subset S^2$, with $S^2$ the unit sphere, we denote the set of direction vectors $\sray$ from the source. For the stereographic projection $\bm{x}$ of this set we use $\Xc\subset\mathbb{R}^2$. The energy density of the source is given by $f=f(\sray)$. We write $\Tc\subset\mathbb{R}^2$ for the set of target positions $\bm{y}$ and $\mathcal{Z}\subset\mathbb{R}^2$ for the set of scaled target variables $\bm{z}$. The desired illuminance on the target plane is then given by $g=g(\bm{y})$ for $\bm{y}\in\Tc$. The energy of any subset of the source domain should be conserved through the optical system. For any subset $\tilde{\mathcal{A}}\subset\Sc$ there is a corresponding set $\mathcal{A}$ of stereographic coordinates. Energy conservation is given by the equation
\begin{equation}\label{eq:base cons}
  \iint_{\tilde{\mathcal{A}}}f(\sray)\,\diff S(\sray) = \iint_{\beta\bm{m}(\mathcal{A})}g(\bm{y})\,\diff A(\bm{y}),
\end{equation}
where $\bm{y}=\beta\bm{z}$. The first integral is over a surface element on the unit sphere, while the second one is over an area element in $\mathbb{R}^2$. When $\mathcal{A}=\Sc$ we have the special case of global energy conservation. This means that the total energy in the source and target should be equal. We assume that $f$ and $g$ are constructed such that this is true. First, we want to transform the left-hand side integral to an integral over $\bm{x}$ instead of $\sray$. From integration by substitution we have
\begin{equation}\label{eq:energy source}
  \iint_{\tilde{\mathcal{A}}}f(\sray)\,\diff S(\sray) = \iint_{\mathcal{A}}f\big(\sray(\bm{x})\big)\left|\frac{\partial\sray}{\partial x_1}\times\frac{\partial\sray}{\partial x_2}\right|\,\diff A(\bm{x}) =\iint_{\mathcal{A}}f\big(\sray(\bm{x})\big)\frac{4}{(|\bm{x}|^2+1)^2}\,\diff A(\bm{x}).
\end{equation}
Now, for the right-hand side of Eq. (\ref{eq:base cons}) we also want to transform the integration variable to $\bm{x}$. For that we first have to change from $\bm{y}$ to $\bm{z}$, since we know $\bm{z}=\bm{m}(\bm{x})$. We had defined $\bm{y}=\beta\bm{z}$, so the Jacobi matrix of this transformation is given by $\beta^2$. Substituting this and Eq. (\ref{eq:energy source}) into Eq. (\ref{eq:base cons}) gives
\begin{equation}\label{eq:energy conserve}
  \iint_{\mathcal{A}}f\left(\sray(\bm{x})\right)\frac{4}{(|\bm{x}|^2+1)^2}\,\diff A(\bm{x}) = \iint_{\mathcal{A}}g\left(\beta\bm{m}(\bm{x})\right)\beta^2|\det(\Diff\bm{m})|\,\diff A(\bm{x}).
\end{equation}
We use $\Diff\bm{m}$ to denote the Jacobi matrix of $\bm{m}$. This equation should hold for any $\mathcal{A}$ corresponding to an $\tilde{\mathcal{A}}\subset\Sc$. Therefore the integrands should be equal almost everywhere. Then for any $\bm{x}(\sray)$ with $\sray\in\Sc$ we have the generalized Monge-Amp\`ere equation
\begin{subequations}\label{eq:Monge-Ampere}
\begin{equation}\label{eq:Monge-Ampere-a}
  |\det(\Diff\bm{m})| =  \frac{4}{\beta^2(|\bm{x}|^2+1)^2}\frac{f\left(\sray(\bm{x})\right)}{g\left(\beta\bm{m}(\bm{x})\right)} =: F(\bm{x},\bm{m}(\bm{x});\beta).
\end{equation}
The boundary condition of the problem is a transport boundary condition given by
\begin{equation}\label{eq:Monge-Ampere-bound}
  \partial\Zc = \bm{m}(\partial\Xc).
\end{equation}
\end{subequations}
It states that the boundary of the source domain should be mapped to the boundary of the target domain \cite{ries94}. For the remainder of this article we will assume that $\det(\Diff\bm{m})>0$, so we can ignore the absolute value in Eq. (\ref{eq:Monge-Ampere-a}).

\subsection{Polar coordinates}
In general, the rays will be emitted by the source in a conical bundle symmetric around the $z$-axis. In that case, the source domain in stereographic coordinates, $\Xc$, will be a circle. So, it makes sense to switch to the polar coordinate system. The polar stereographic coordinates are written as $\bm{\omega}=(\rho,\zeta)$ and the transformation $\bm{x}=\bm{x}(\bm{\omega})$ is given by
\begin{equation}
  x_1 = \rho\cos\zeta ,\qquad x_2 = \rho\sin\zeta.
\end{equation}
We define $\Omega$ to be the source domain in polar stereographic coordinates, so $\Omega=\bom(\Xc)$. Furthermore we define $u_1^*(\bm{\omega})=u_1(\bm{x}(\bm{\omega}))$ and $c^*(\bm{\omega},\bm{z})=c(\bm{x}(\bm{\omega}),\bm{z})$, so equation (\ref{eq:main1}) changes to
\begin{equation}\label{eq:main polar}
  u_1^*(\bm{\omega})+u_2(\bm{z}) = c^*(\bm{\omega},\bm{z}).
\end{equation}
We also need to transform the energy conservation equation to polar coordinates. With integration by substitution, we change Eq. (\ref{eq:energy conserve}) to
\begin{equation}\label{eq:energy polar}
  \iint_{\bm{\omega}(\mathcal{A})}f\left(\sray(\bom)\right)\frac{4}{(\rho^2+1)^2}\rho\,\diff\bom = \iint_{\bm{\omega}(\mathcal{A})}g\left(\beta\bm{m}^*(\bom)\right)\beta^2\det(\Diff\bm{m}^*)\rho\,\diff\bom,
\end{equation}
where $\diff\bom=\diff\rho\diff\zeta$. We introduce the notation $\bm{m}^*(\bm{\omega})=\bm{m}\left(\bm{x}(\bom)\right)$, implying $\bm{z}=\bm{m}^*(\bom)$. Note that Eq. (\ref{eq:energy polar}) should hold for $\mathcal{A}$ corresponding to any arbitrary part of the source domain, so for any $\bm{\omega}\in\Omega$ we have
\begin{subequations}\label{eq:Monge polar}
\begin{equation}\label{eq:Monge polar-a}
  \det(\Diff \bm{m}^*) = \frac{4}{\beta^2(\rho^2+1)^2}\frac{f\left(\sray(\bom)\right)}{g\left(\beta\bm{m}^*(\bom)\right)} := F^*\left(\bom,\bm{m}^*(\bom)\right).
\end{equation}
The matrix $\Diff\bm{m}^*$ is the Jacobian of $\bm{m}$ expressed in polar coordinates and is given by
\begin{equation}
  \Diff\bm{m}^* = \begin{pmatrix}
                    \frac{\partial m_1}{\partial\rho} & \frac{1}{\rho}\frac{\partial m_1}{\partial\zeta} \\
                    \frac{\partial m_2}{\partial\rho} & \frac{1}{\rho}\frac{\partial m_2}{\partial\zeta}
                  \end{pmatrix}.
\end{equation}
This follows from a coordinate transformation of the Jacobian with Cartesian coordinates in Section \ref{sec:energy}. The boundary condition (\ref{eq:Monge-Ampere-bound}) changes to
\begin{equation}\label{eq:Monge polar boundary}
  \partial\Zc = \bm{m}^*\left(\bom\left(\partial\Xc\right)\right).
\end{equation}
\end{subequations}
Note that we have $\Omega=\bom(\Xc)$, but $\partial\Omega\neq\bom\left(\partial\Xc\right)$. For example, the relation $\rho=0$ would give a point in the interior of $\Xc$, but a line on the boundary of $\Omega$. In the remainder of this article we will omit the asterisks, because we will only work with functions of $\bom$.

\subsection{Mapping}\label{sec:mapping}
We want to find an (implicit) expression for the optical mapping $\bm{z}=\bm{m}(\bom)$. This can be done using Eq. (\ref{eq:main polar}). There are many solutions to solve this equation for $\bm{m}$, so we make a special choice \cite{yad18}. It is possible to find a $c$-convex pair of functions $u_1$, $u_2$ such that
\begin{equation}
  u_1(\bom) = \max_{\bm{z}\in\Zc}\left(c(\bom,\bm{z})-u_2(\bm{z})\right), \quad u_2(\bm{z}) = \max_{\bom\in\Omega}\left(c(\bom,\bm{z})-u_1(\bom)\right).
\end{equation}
Conversely it is possible to find a $c$-concave pair defined by
\begin{equation}
  u_1(\bom) = \min_{\bm{z}\in\Zc}\left(c(\bom,\bm{z})-u_2(\bm{z})\right), \quad u_2(\bm{z}) = \min_{\bom\in\Omega}\left(c(\bom,\bm{z})-u_1(\bom)\right).
\end{equation}
In either case, the solution for $u_2$ has an argument $\hat{\bom}$ that is a stationary point. This leads to the requirement that
\begin{equation}\label{eq:stationary}
  \nabla_{\bom}c(\hat{\bom},\bm{z})-\nabla_{\bom} u_1(\hat{\bom}) = \bm{0},
\end{equation}
where the gradient with respect to $\bom$ is given by
\begin{equation}
  \nabla_{\bom} = \bm{e}_\rho\frac{\partial}{\partial\rho}+ \frac{1}{\rho}\bm{e}_\zeta\frac{\partial}{\partial\zeta}.
\end{equation}
To ensure that the solution is a maximum we need the Hessian w.r.t. $\bom$ of $c(\bom,\bm{z})-u_1(\bom)$ to be symmetric negative definite (SND). Similarly, to ensure a minimum we require the Hessian to be symmetric positive definite (SPD). The Hessian matrix of any function $v$ in polar coordinates is given by \cite{bel15}
\begin{equation}
  H[v] = \begin{pmatrix}
           \frac{\partial^2 v}{\partial\rho^2} & \frac{1}{\rho}\frac{\partial^2v}{\partial\rho\partial\zeta}-\frac{1}{\rho^2}\frac{\partial v}{\partial\zeta} \\
           \frac{1}{\rho}\frac{\partial^2 v}{\partial\rho\partial\zeta}-\frac{1}{\rho^2}\frac{\partial v}{\partial\zeta} & \frac{1}{\rho^2}\frac{\partial^2 v}{\partial\zeta^2}+\frac{1}{\rho}\frac{\partial v}{\partial\rho}
         \end{pmatrix}.
\end{equation}
In the Hessian of the function $c(\bom,\bm{z})-u_1(\bom)$ the first derivative terms cancel because of Eq. (\ref{eq:stationary}), so this Hessian is given by
\begin{equation}\label{eq:Hessian}
  H\left[c(\bom,\bm{z})-u_1(\bom)\right] = \Diff_{\bom\bom}c(\bom,\bm{z})-\Diff_{\bom\bom}u_1(\bom),
\end{equation}
with
\begin{equation}
  \Diff_{\bm{\omega\omega}} = \begin{pmatrix}
                                \frac{\partial^2}{\partial\rho^2} & \frac{1}{\rho}\frac{\partial^2}{\partial\rho\partial\zeta} \\
                                \frac{1}{\rho}\frac{\partial^2}{\partial\zeta\partial\rho} & \frac{1}{\rho^2}\frac{\partial^2}{\partial\zeta^2}
                              \end{pmatrix}.
\end{equation}
Note that $\Diff_{\bom\bom}v$ is not equal to the Hessian matrix of $v$. We assume that a mapping $\bm{z}=\bm{m}(\bom)$ exists, and substitute this into Eq. (\ref{eq:stationary}). We then take the derivative with respect to $\bom$ and apply the chain rule to obtain
\begin{subequations}
\begin{equation}\label{eq:stationary diff}
  \Diff_{\bom\bom}c+\bm{C}\Diff\bm{m} = \Diff^2u_1,
\end{equation}
where
\begin{equation}
\bm{C}=\Diff_{\bom\bm{z}}c = \begin{pmatrix}
                            \frac{\partial^2 c}{\partial\rho\partial z_1} & \frac{\partial^2 c}{\partial\rho\partial z_2} \\
                            \frac{1}{\rho}\frac{\partial^2 c}{\partial\zeta\partial z_1} & \frac{1}{\rho}\frac{\partial^2 c}{\partial\zeta\partial z_2}
                          \end{pmatrix}.
\end{equation}
\end{subequations}
Note that the first term in Eq. (\ref{eq:stationary diff}) means differentiating $c(\bom,\bm{z})$ w.r.t. $\bom$ twice and then substituting $\bm{z}=\bm{m}(\bom)$. We can rewrite Eq. (\ref{eq:stationary diff}) to
\begin{equation}\label{eq:cdm}
  \bm{C}\Diff\bm{m} = \Diff_{\bom\bom}u_1-\Diff_{\bom\bom}c =: \bm{P}.
\end{equation}
Note that $-\bm{P}$ is the Hessian matrix in Eq. (\ref{eq:Hessian}). Therefore, for a $c$-convex or $c$-concave pair of functions $u_1$, $u_2$, we have the condition that $\bm{P}$ should be SPD or SND, respectively. We assumed that a mapping $\bm{m}$ is defined (implicitly) by Eq. (\ref{eq:stationary}). By the implicit function theorem such a mapping is guaranteed to exist if the Jacobian matrix of the left-hand side of the equation with respect to $\bm{z}$ is invertible \cite[Sec.~12.8]{ada13}. \\
\indent To summarize, we need to find a mapping $\bm{m}$ satisfying Eq. (\ref{eq:Monge polar boundary}) and Eq. (\ref{eq:cdm}) for a matrix $\bm{P}$ that is SPD or SND, with $\det(\bm{P})=F\det(\bm{C})$. The exact mapping is determined by the parameters $l$ and $\beta$ (or $L$) and the direction of the outgoing beam, given by $\varphi$ and $\theta$.

\section{Numerical method}\label{sec:algorithm}
We explain the least-squares algorithm which we use to solve the problem derived in the previous section. This algorithm has been explained thoroughly for a source with Cartesian \cite{prins15,yad18} and polar \cite{rom19} coordinates. In this section we give a brief overview. We first compute the mapping, followed by a calculation of the surfaces. We restrict ourselves to the $c$-convex solution of Eq. (\ref{eq:main polar}). As shown in the previous section, we need to solve
\begin{equation}\label{eq:CDP}
  \bm{C}(\bm{\omega},\bm{m}(\bm{\omega}))\Diff\bm{m}(\bm{\omega}) = \bm{P}(\bm{\omega}),
\end{equation}
where $\bm{C}=\Diff_{\bom\bm{z}}c$ and $\bm{P}$ satisfies $\det(\bm{P})=F\det(\bm{C})$, see Eq. (\ref{eq:Monge polar-a}). To get the $c$-convex solution, $\bm{P}$ needs to be SPD. We enforce the equality in Eq (\ref{eq:CDP}) by minimizing the functional $J_\text{I}$ defined by
\begin{equation}
  J_\text{I}[\bm{m},\bm{P}] = \frac{1}{2}\iint_{\Omega}\|\bm{C}\Diff\bm{m}-\bm{P}\|_\text{F}^2\rho\,\diff\bom.
\end{equation}
The norm $\|.\|_\text{F}$ is the Frobenius norm. To enforce the boundary condition we minimize the difference between the (given) boundary of $\Tc$ and the mapping of the boundary of the source:
\begin{equation}
  J_\text{B}[\bm{m},\bm{b}] = \frac{1}{2}\int_{\bom(\partial\Xc)}|\bm{m}-\bm{b}|^2\,\diff s,
\end{equation}
where $\bm{b}\in\partial\Zc$. We combine these two functionals into a weighted average with parameter $\alpha\in[0,1]$ given by
\begin{equation}\label{eq:functional mix}
  J[\bm{m},\bm{P},\bm{b}] = \alpha J_\text{I}[\bm{m},\bm{P}] + (1-\alpha)J_\text{B}[\bm{m},\bm{b}].
\end{equation}
All of these functionals are defined on the following spaces
\begin{subequations}
  \begin{align}
    \mathcal{P}(\bm{m}) &= \left\{\bm{P}\in[C^1(\Omega)]^{2\times2} \mid \det(\bm{P})=F\det(\bm{C}),\ \bm{P}\text{ SPD}\right\}, \\
    \mathcal{B} &= \left\{\bm{b}\in[C^1(\bom(\partial\Xc))]^2 \mid \bm{b}(\bom)\in\partial\Zc\right\}, \\
    \mathcal{M} &= [C^2(\Omega)]^2 .
  \end{align}
\end{subequations}
We cover the domain $\Omega$ by a grid with gridpoints $\bom_{ij}$. The algorithm to find $\bm{m}$ is initialized by a guess $\bm{m}^0$ for the mapping. With this mapping we compute the matrix $\bm{C}^0$. Then, we iteratively perform the next steps either for a fixed number of iterations or until a stopping criterion is met,
\begin{subequations}\label{eq:algorithm}
  \begin{align}
    \bm{b}^{n+1} &= \argmin_{\bm{b}\in\mathcal{B}}J_\text{B}[\bm{m}^n,\bm{b}], \label{eq:min b}\\
    \bm{P}^{n+1} &= \argmin_{\bm{P}\in\mathcal{P}(\bm{m}^n)}J_\text{I}[\bm{m}^n,\bm{P}], \label{eq:min P}\\
    \bm{m}^{n+1} &= \argmin_{\bm{m}\in\mathcal{M}}J[\bm{m},\bm{P}^{n+1},\bm{b}^{n+1}], \label{eq:min m}\\
    \bm{C}^{n+1} &= \bm{C}(\bom,\bm{m}^{n+1}).
  \end{align}
\end{subequations}
The minimization procedures for $\bm{b}$ and $\bm{P}$ do not contain derivatives of their respective variables, so these can be minimized pointwise. \\
\indent A method for solving step (\ref{eq:min b}) is given by Romijn et al. \cite{rom20}. We will explain step (\ref{eq:min P}) in a bit more detail. The matrix $\bm{P}$ needs to be SPD. The symmetry of $\bm{P}$ is enforced by defining
\begin{equation}
  \bm{P} = \begin{pmatrix}
             p_{11} & p_{12} \\
             p_{12} & p_{22}
           \end{pmatrix}.
\end{equation}
We approximate $\Diff\bm{m}$ using central differences and define $\bm{Q}=\bm{C}\Diff\bm{m}$. Instead of minimizing $\|\bm{Q}-\bm{P}\|_\text{F}$ we solve an equivalent problem with the same minimizers \cite{prins15}. We introduce the symmetric matrix $\bm{Q}_\text{S}=\frac{1}{2}(\bm{Q}+\bm{Q}\tran)$, with off-diagonal entries $q_\text{S}=\frac{1}{2}(q_{12}+q_{21})$. The optimization problem that needs to be solved is then
\begin{equation}\label{eq:min sym}
  \begin{aligned}
  \text{minimize}\quad&H_\text{S}(p_{11},p_{22},p_{12})=\frac{1}{2}\|\bm{Q}_\text{S}-\bm{P}\|_\text{F}^2, \\
  \text{subject to}\quad&\det(\bm{P}) = F\det(\bm{C}).
  \end{aligned}
\end{equation}
It turns out that we can always select at least one solution of this problem that satisfies the constraint that $\bm{P}$ is SPD \cite{yad18}. To solve problem (\ref{eq:min sym}) we use the Lagrange multiplier method. We introduce the Lagrangian function
\begin{equation}
  \Lambda(p_{11},p_{22},p_{12},\lambda) = H_\text{S}(p_{11},p_{22},p_{12})+\lambda\left(p_{11}p_{22}-p_{12}^2-F\det(\bm{C})\right).
\end{equation}
To find stationary points we take the derivatives w.r.t. each variable and set them equal to zero. Elementary calculation lead us then to the system of equations
\begin{subequations}
  \begin{align}
    p_{11}+\lambda p_{22} &= q_{11}, \\
    \lambda p_{11}+p_{22} &= q_{22}, \\
    (1-\lambda)p_{12} &= q_\text{S}, \\
    p_{11}p_{22}-p_{12}^2 &= F\det(\bm{C}).
  \end{align}
\end{subequations}
Solutions of this system can be calculated analytically and explicitly\cite{yad18}. In the case that we find multiple solutions, we have to choose the one that gives the lowest value for $H_\text{S}$.

\subsection*{Computing the mapping}
In the functional $J$ there are derivatives of $\bm{m}$, so we can no longer optimize pointwise. To be able to minimize this functional, we apply calculus of variations. The first variation of $\bm{m}$ in the direction of an arbitrary function $\bm{\eta}=(\eta_1,\eta_2)\tran\in\mathcal{M}$ is given by
\begin{equation}\label{eq:variation1}
\begin{aligned}
  \delta J[\bm{m},&\bm{P},\bm{b}](\bm{\eta}) = \lim_{\varepsilon\to0}\frac{1}{\varepsilon}\bigg(J[\bm{m}+\varepsilon\bm{\eta},\bm{P},\bm{b}]-J[\bm{m},\bm{P},\bm{b}]\bigg) \\
  &=\lim_{\varepsilon\to0}\frac{1}{\varepsilon}\left(\frac{\alpha}{2}\iint_{\Omega}\|\bm{C}\Diff\bm{m}+\varepsilon\bm{C}\Diff\bm{\eta}-\bm{P}\|_\text{F}^2\rho\,\diff\bom +\frac{1-\alpha}{2}\int_{\bom(\partial\Xc)}|\bm{m}+\varepsilon\bm{\eta}-\bm{b}|^2\,\diff s\right. \\ &\qquad\left. -\frac{\alpha}{2}\iint_{\Omega}\|\bm{C}\Diff\bm{m}-\bm{P}\|_\text{F}^2\rho\,\diff\bom -\frac{1-\alpha}{2}\int_{\bom(\partial\Xc)}|\bm{m}-\bm{b}|^2\,\diff s\right) \\
  &=\lim_{\varepsilon\to0}\frac{1}{\varepsilon}\left(\alpha\iint_{\Omega}\varepsilon\rho(\bm{C}\Diff\bm{m}-\bm{P})\bm{:}\bm{C}\Diff\bm{\eta} +\frac{\varepsilon^2}{2}\rho\|\bm{C}\Diff\bm{\eta}\|_\text{F}^2\,\diff\bom\right. \\ &\qquad \left. +(1-\alpha)\int_{\bom(\partial\Xc)}\varepsilon(\bm{m}-\bm{b})\bm{\cdot}\bm{\eta} +\frac{\varepsilon^2}{2}|\bm{\eta}|^2\,\diff s\right) \\
  &=\alpha\iint_{\Omega}(\bm{C}\Diff\bm{m}-\bm{P})\bm{:}\bm{C}\Diff\bm{\eta}\,\rho\,\diff\bom +(1-\alpha)\int_{\bom(\partial\Xc)}(\bm{m}-\bm{b})\bm{\cdot}\bm{\eta}\,\diff s,
\end{aligned}
\end{equation}
where $\bm{:}$ denotes the Frobenius inner product associated with the Frobenius norm $\|.\|_F$. Like before, $\Diff\bm{\eta}$ denotes the Jacobian of $\bm{\eta}$ w.r.t. $\bom$. We can rewrite the inner product in the first integral to obtain
\begin{equation}
  (\bm{C}\Diff\bm{m}-\bm{P})\bm{:}\bm{C}\Diff\bm{\eta} = \bm{w}_1\bm{\cdot}\nabla\eta_1+\bm{w}_2\bm{\cdot}\nabla\eta_2,
\end{equation}
where $\bm{w}_1$ and $\bm{w}_2$ are defined by
\begin{equation}\label{eq:def w}
  \begin{bmatrix}
    \bm{w}_1\tran \\
    \bm{w}_2\tran
  \end{bmatrix} = \bm{C}\tran(\bm{C}\Diff\bm{m}-\bm{P}).
\end{equation}
We use this and Gauss's divergence theorem to rewrite the integral over $\Omega$ as
\begin{equation}
  \begin{aligned}
  \iint_{\Omega}(\bm{C}\Diff\bm{m}-\bm{P})\bm{:}\bm{C}\Diff\bm{\eta}\,\rho\,\diff\bom &=\sum_{k=1}^2\iint_{\Omega}\bm{w}_k\bm{\cdot}\nabla\eta_k\,\rho\,\diff\bom \\
  &=\sum_{k=1}^2\left[\oint_{\bom(\partial\Xc)}\eta_k\bm{w}_k\bm{\cdot}\nvec\,\diff s-\iint_{\Omega}\eta_k\diver(\bm{w}_k)\,\rho\,\diff\bom\right],
  \end{aligned}
\end{equation}
where $\nvec$ is the outward unit normal of $\Xc$. We substitute this into Eq. (\ref{eq:variation1}) and set the first variation equal to zero to obtain
\begin{equation}
  \sum_{k=1}^2\left[\oint_{\bom(\partial\Xc)}\eta_k\left(\alpha\bm{w}_k\bm{\cdot}\nvec+(1-\alpha)(m_k-b_k)\right)\,\diff s-\iint_{\Omega}\alpha\eta_k\diver(\bm{w}_k)\,\rho\,\diff\bom\right] = 0.
\end{equation}
At a minimum of $J$, the first variation should be equal to zero for any vector $\bm{\eta}$. We can split this in two cases. First we choose $\eta_2=0$ and set the first variation equal to zero for any $\eta_1\in C^2(\Omega)$. Similarly, we have the case where $\eta_1=0$. With the use of the fundamental lemma of calculus of variations \cite{cou08} we get the boundary value problem
\begin{subequations}
  \begin{align}
    &\diver(\bm{C}^T\bm{C}\Diff\bm{m}) = \diver(\bm{C}^T\bm{P}), \quad&&\text{for }\bom\in\Omega, \label{eq:internal}\\
    &\alpha\bm{C}^T\bm{C}(\Diff\bm{m})\nvec+(1-\alpha)\bm{m} = \alpha\bm{C}^T\bm{P}\nvec+(1-\alpha)\bm{b}, \quad&&\text{for }\bom\in\bom(\partial\Xc), \label{eq:minM boundary}
  \end{align}
\end{subequations}
where $\diver$ is defined in the following way. Let $\bm{B}=(b_{ij})\in\mathbb{R}^{2\times2}$, then
\begin{equation}
  \diver(\bm{B}) = \frac{1}{\rho}\begin{pmatrix}
                       \frac{\partial}{\partial\rho}(\rho b_{11})+\frac{\partial b_{12}}{\partial\zeta} \\
                       \frac{\partial}{\partial\rho}(\rho b_{21})+\frac{\partial b_{22}}{\partial\zeta}
                     \end{pmatrix}.
\end{equation}
This boundary value problem is then solved using the finite volume method \cite[App.~A]{rom19}.

\subsection*{Computation of the reflector surfaces}\label{sec:surfaces}
The algorithm (\ref{eq:algorithm}) computes a mapping $\bm{m}$. From this mapping we can compute the shape of the surfaces, given by $u$ and $w$. To find $u$ we first compute $u_1$ from equation (\ref{eq:stationary}). We introduce the functional $I$ to quantify how close a function $\phi$ is to the exact solution of that equation. Our solution $u_1$ is then given by
\begin{equation}
  u_1 = \argmin_\phi I[\phi],
\end{equation}
where
\begin{equation}\label{eq:functional reflector}
  I[\phi] = \frac{1}{2}\iint_\Omega|\nabla\phi-\nabla_{\bom} c(\cdot\, ,\bm{m})|^2\rho\,\diff\bom.
\end{equation}
To solve this optimization problem we use calculus of variations. The first variation of (\ref{eq:functional reflector}) is
\begin{equation}
  \delta I[u_1](v) = \frac{1}{2}\iint_\Omega(\nabla u_1-\nabla_{\bom} c)\bm{\cdot}\nabla v\,\rho\,\diff\bom,
\end{equation}
analogous to Eq. (\ref{eq:variation1}). Similar to the calculation of the mapping, we use Gauss's divergence theorem and the fundamental lemma of calculus of variations to get the boundary value problem
\begin{subequations}\label{eq:bvp refl}
  \begin{align}
    &\Delta u_1 = \diver(\nabla_{\bom}c),\qquad&&\bom\in\Omega \\
    &\frac{\partial u_1}{\partial\rho} = \left.\frac{\partial c(\cdot\, ,\bm{z})}{\partial\rho}\right|_{\bm{z}=\bm{m}(\bom)}, \qquad&&\bom\in\bom\left(\partial\Xc\right).
  \end{align}
\end{subequations}
Here, $\Delta$ denotes the Laplace operator, which in polar coordinates is given by
\begin{equation}
  \Delta u_1 = \diver(\nabla u_1) = \frac{1}{\rho}\frac{\partial}{\partial\rho}\left(\rho\frac{\partial u_1}{\partial\rho}\right)+\frac{1}{\rho^2}\frac{\partial^2u_1}{\partial\zeta^2}.
\end{equation}
Let $u_1$ be a solution to this boundary value problem. To calculate the shapes of the reflector surfaces, $u$ and $w$, we combine Eq. (\ref{eq:kappa defs}) and Eq. (\ref{eq:shapes}) to obtain
\begin{equation}\label{eq:shape calculation}
  u(\sray) = \frac{\beta}{1-\sray\bm{\cdot}\bm{\hat{a}}_3+e^{u_1}},\qquad w(\bm{z}) = l+\frac{1}{2}\beta\left(1-|\bm{z}|^2\right)-\beta e^{u_2},
\end{equation}
where $u_2=c-u_1$, using Eq. (\ref{eq:main1}). \\
\indent The solution to the boundary value problem (\ref{eq:bvp refl}) is unique up to an additive constant \cite{rom19}. This means that choosing the value of $u_1$ in one point gives us a unique solution. We use this degree of freedom to choose the position of one of the reflectors along the central ray given by $\sray=\sray_0:=(0,0,1)\tran$. For example, we can choose $u(\sray_0)=u_0$ for some $u_0>0$. With Eq. (\ref{eq:shape calculation}), this then gives
\begin{equation}
  u_1(\bom_0) = \log\bigg(\frac{\beta}{u_0}-(1-\sray_0\bm{\cdot}\bm{\hat{a}}_3)\bigg) = \log\bigg(\frac{\beta}{u_0}-(1-\cos\varphi)\bigg),
\end{equation}
where $\bom_0=\bom(\sray_0)$ and $\varphi$ is the polar angle of the output rays. This expression, together with the boundary value problem (\ref{eq:bvp refl}), gives a unique solution for $u_1$ and therefore also for $u$ and $w$. In this case we cannot freely choose the position of the second reflector along the central ray. Alternatively, we could give $w\big(\bm{m}(\bom_0)\big)=w_0$ as input instead of $u_0$. This would also fix $u_1$, $u$ and $w$.

\section{Results}
We will apply the method from the previous section to examples with two distinct combinations of source and target distributions. We create several optical systems with different layouts to show the possibilities of the algorithm.

\subsection{Uniform to uniform}
\begin{figure}[htbp]
  \centering
  \includegraphics[width=0.45\linewidth]{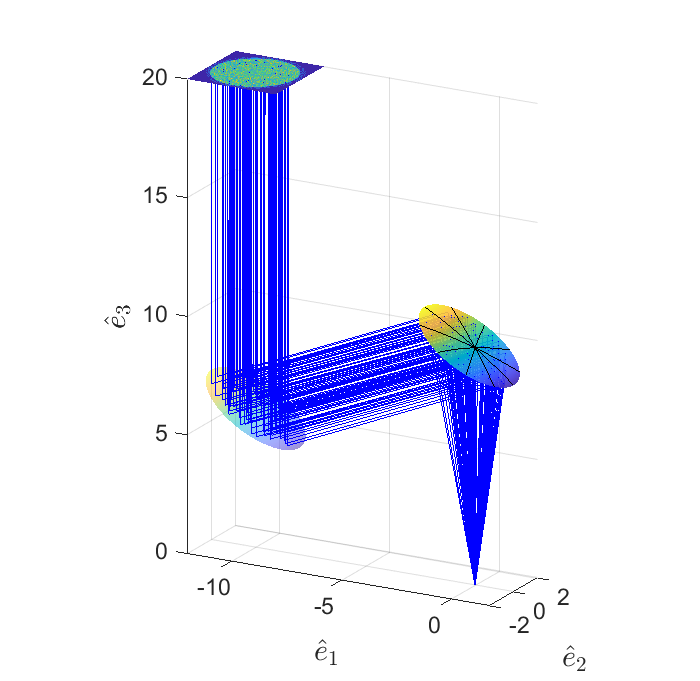}
  \caption{The resulting optical system for the first test case with $\varphi=0$, $u_0=10$ and $\beta=15$.}\label{fig:test1 overview}
\end{figure}

\begin{figure}[htbp]
  \centering
  \begin{subfigure}{0.45\textwidth}
    \centering
    \includegraphics[width=\textwidth]{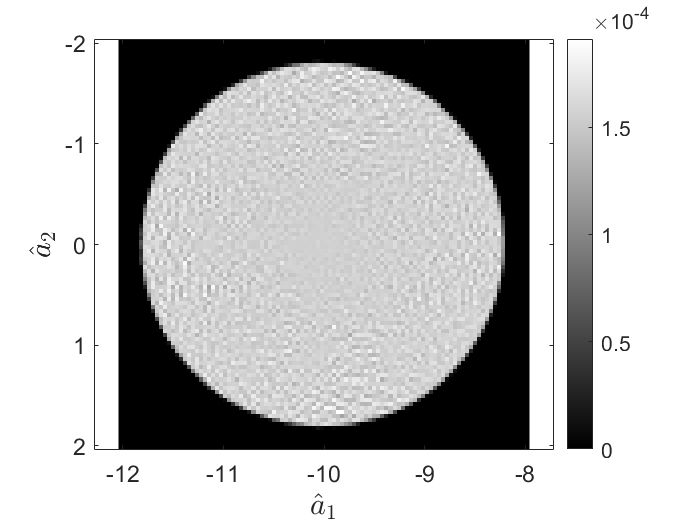}
    \caption{Illumination pattern}
    \label{fig:test1 pattern}
  \end{subfigure}
  \begin{subfigure}{0.45\textwidth}
    \centering
    \includegraphics[width=\textwidth]{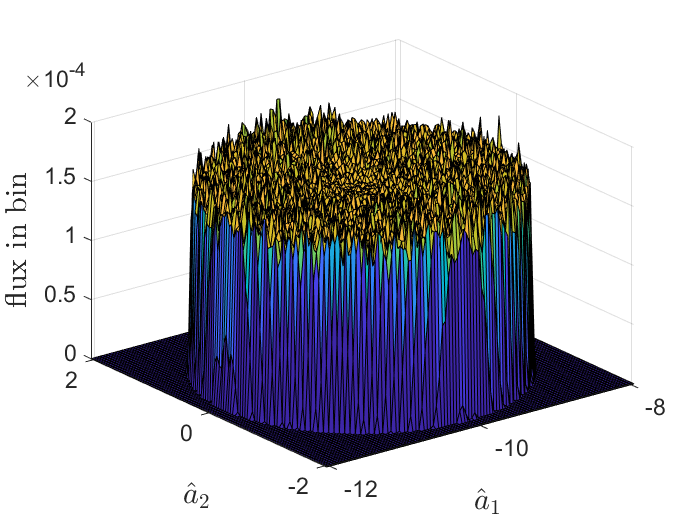}
    \caption{3-dimensional flux plot}
    \label{fig:test1 3d}
  \end{subfigure}
  \caption{The ray traced light flux per bin on the target domain visualized in two ways.}\label{fig:test1 illumination}
\end{figure}

\begin{figure}[htbp]
  \centering
  \begin{subfigure}{0.45\textwidth}
    \centering
    \includegraphics[width=\textwidth]{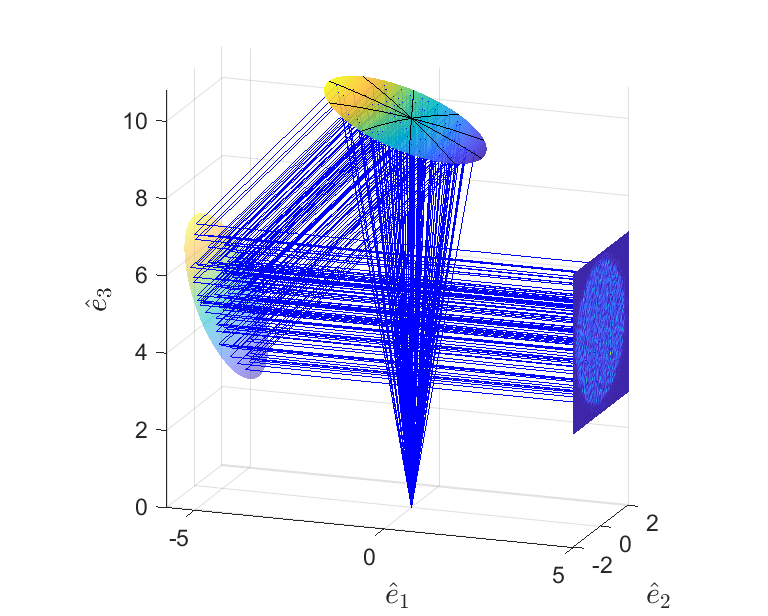}
    \caption{$\varphi=\frac{\pi}{2}$, $\beta=22$, $u_0=10$.}
    \label{fig:layouts1}
  \end{subfigure}
  \begin{subfigure}{0.45\textwidth}
    \centering
    \includegraphics[width=\textwidth]{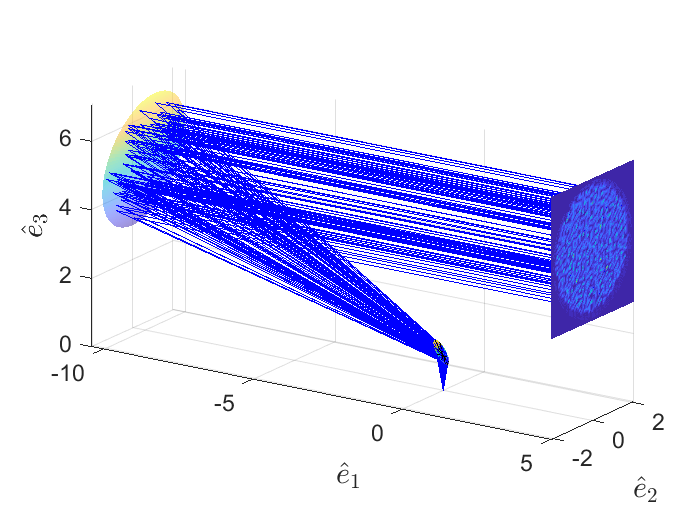}
    \caption{$\varphi=\frac{\pi}{2}$, $\beta=22$, $u_0=1$.}
    \label{fig:layouts2}
  \end{subfigure}
  \begin{subfigure}{0.45\textwidth}
    \centering
    \includegraphics[width=\textwidth]{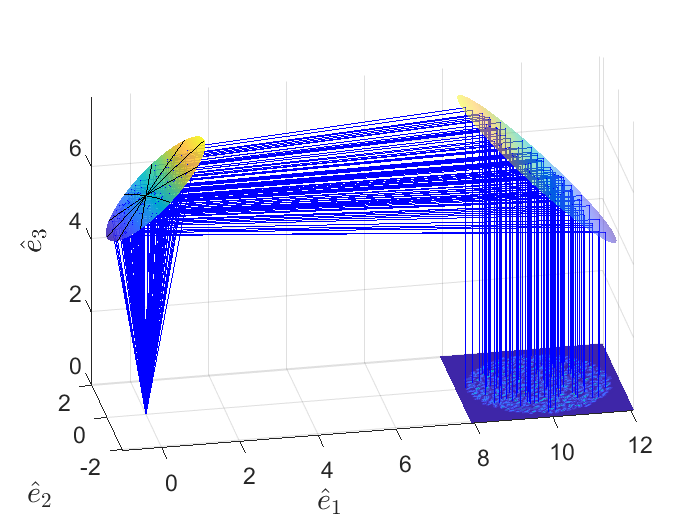}
    \caption{$\varphi=\pi$, $\beta=22$, $w_0=6$.}
    \label{fig:layouts3}
  \end{subfigure}
  \begin{subfigure}{0.45\textwidth}
    \centering
    \includegraphics[width=\textwidth]{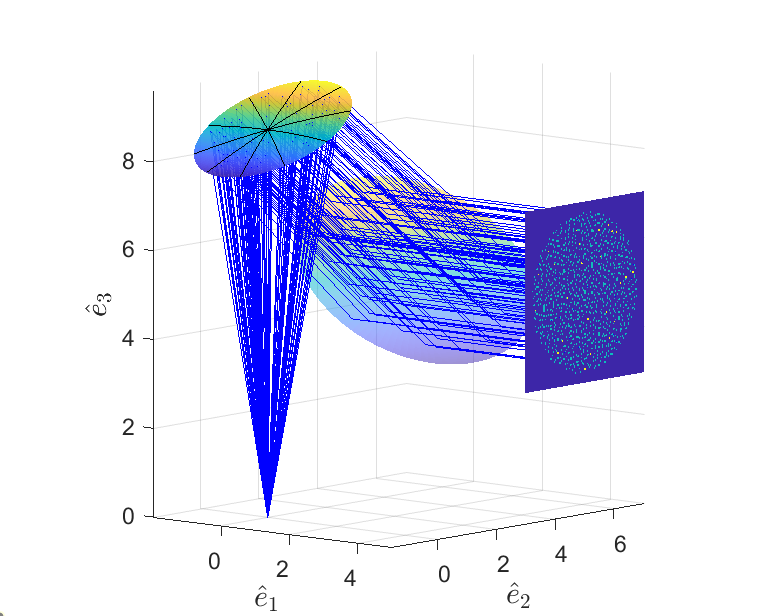}
    \caption{$\varphi=\frac{\pi}{2}$, $\beta=15$, $w_0=5$.}
    \label{fig:layouts4}
  \end{subfigure}
  \caption{Some examples of layouts constructed with our algorithm}\label{fig:layouts}
\end{figure}

The example we will look at first consists of a uniform conical source and a uniform circular target distribution. We will test several layouts of optical systems, so different angles for the outgoing beam. Because our source is rotationally symmetric, our choice of $\theta$ does not matter and we can choose $\theta=0$. We will discuss optical systems with $\varphi=0$, $\varphi=\pi/2$ and $\varphi=\pi$. The first and last one have outgoing beams parallel to the $\bm{\hat{e}}_3$-axis in the positive and negative direction, respectively. For $\varphi=\pi/2$ the output beam is parallel to the $\bm{\hat{e}}_1$-axis. We shift the target domain along the target plane to avoid mirrors obstructing rays. Otherwise, for example, the first reflector could be in the way of the rays from the second reflector to the target. For the point source, we use a uniform distribution in the direction vector $\sray$ rather than the stereographic coordinates $\bm{x}$. This means that the intensity distribution given by $f=f(\sray)$ is constant. The source domain is given by $\rho\leq0.1$ and we choose the value of $f$ such that the total flux is $1$. The target domain is a circle with radius 2. The target distribution $g$ is constant over this domain, with a flux of $1$. \\
\indent There are some parameter choices that will affect the optimization procedure or the resulting optical system. For example, the functional in equation (\ref{eq:functional mix}) contains the parameter $\alpha$. The smaller $\alpha$, the more important the boundary is in the optimization, relative to the interior. The choice of $\alpha$ will have an impact on the convergence speed of the algorithm \cite{yad18}. Other parameters have an influence on the layout of the optical system. These are the distance to the target plane, $l$, the reduced optical path length, $\beta$ (or $L$), and either $u_0$ or $w_0$. \\
\indent For our first test case we use $\alpha=0.1$. From numerical experiments we found that this value works well. We discretize the polar source domain with a $200\times200$ grid. The results of our test case were obtained by running the least-squares algorithm for 200 iterations. First, we did this for a case where $\varphi=0$. We use a target plane at a distance $l=20$ and we choose to shift the target domain by a distance $-10$ along the $\bm{\hat{a}}_1$-direction. Furthermore, we put $u_0=10$ and $\beta=15$. The resulting optical system is visualized using a ray trace procedure. The result can be seen in Fig. \ref{fig:test1 overview}.
In the figure we show 100 random rays that are obtained from this ray trace. On the target plane we show the illumination pattern in a bounding box of the target domain. We divide this box in 100 by 100 bins and use a quasi-Monte Carlo method tracing ten million rays to get an illumination pattern. A better view of the illumination pattern is given in Fig. \ref{fig:test1 illumination}.
As we can see, the illuminated part of the bounding box forms a circle. The flux per bin, and thus the intensity, on this circle is nicely uniform. The bins within the circle have some variation in flux that is caused by the Monte Carlo method. \\
\indent As mentioned before, an important benefit of our algorithm is the ability to construct a wide variety of optical systems. Some examples of possible layouts are shown in Fig. \ref{fig:layouts}. The first two figures both have $\varphi=\pi/2$, $\theta=0$ and $\beta=22$. The target plane is located at a distance $l=5$ and the center of the target domain is shifted by $-5$ along $\bm{\hat{a}}_1$. With $\varphi=\pi/2$ and $\theta=0$ we have $\bm{\hat{a}}_1=-\bm{\hat{e}}_3$. The difference between the two optical systems is that in Fig. \ref{fig:layouts1} we have $u_0=10$ and in Fig. \ref{fig:layouts2} we have $u_0=1$. Note that these two have the exact same mapping, since the parameter $u_0$ only plays a role in the calculation of the surfaces (see Sec. \ref{sec:surfaces}). Therefore, we only need to calculate the mapping once, and we are able to construct both of these optical systems from this mapping by varying $u_0$. For Fig. \ref{fig:layouts3} we change the angle of the outgoing beam to $\varphi=\pi$ and we use $w_0=6$. We choose $l=0$, so the target plane is equal to the source plane. We shift the target domain by $10$ along the $\bm{\hat{e}}_1$-axis and we again use $\beta=22$. In each of the previously mentioned optical systems, there is a plane of symmetry. However, this is not necessary for our algorithm. In Fig. \ref{fig:layouts4}, we shift the target domain by $-5$ along $\bm{\hat{a}}_1$ and 5 along $\bm{\hat{a}}_2$. This breaks the symmetry of the optical system. Furthermore, we set $\beta=15$ and $w_0=5$.

\subsection{Laser diode to ring-shaped target}
\begin{figure}[htbp]
  \centering
  \begin{subfigure}{0.4\textwidth}
    \includegraphics[width=\linewidth]{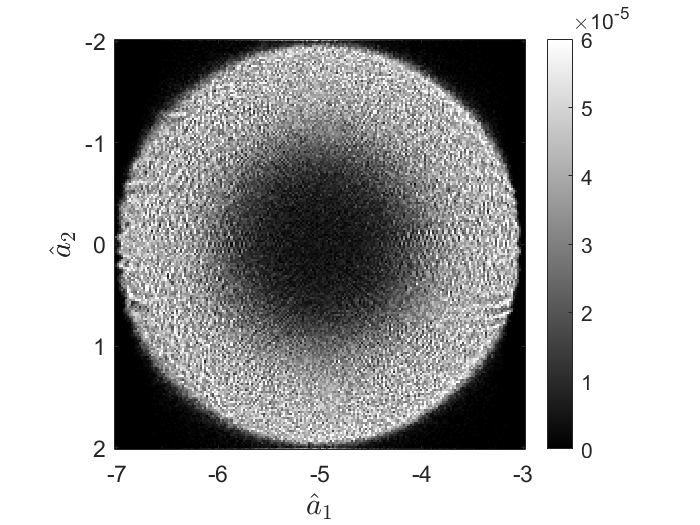}
    \caption{illumination pattern, using $200\times200$ bins.}
  \end{subfigure}
  \hspace{0.05\linewidth}
  \begin{subfigure}{0.4\textwidth}
    \includegraphics[width=\linewidth]{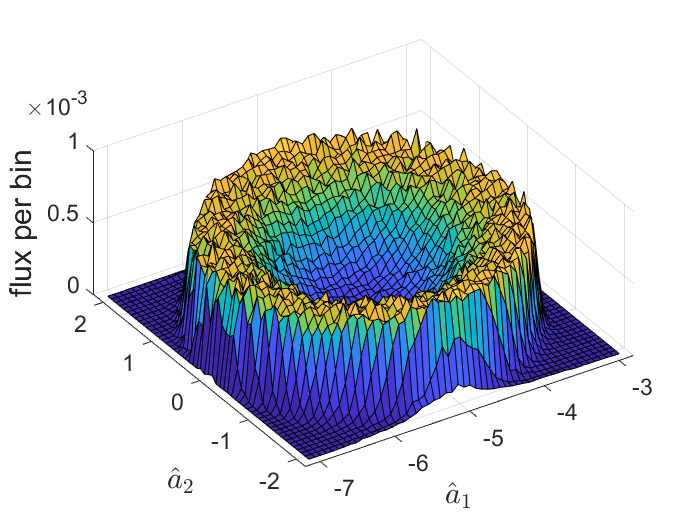}
    \caption{3-dimensional density plot, using $50\times50$ bins.}
  \end{subfigure}
  \caption{The ray traced light intensity on the target plane for a laser diode source and ring-shaped target intensity.}\label{fig:donut illum}
\end{figure}

\begin{figure}[htbp]
  \centering
  \includegraphics[width=\linewidth]{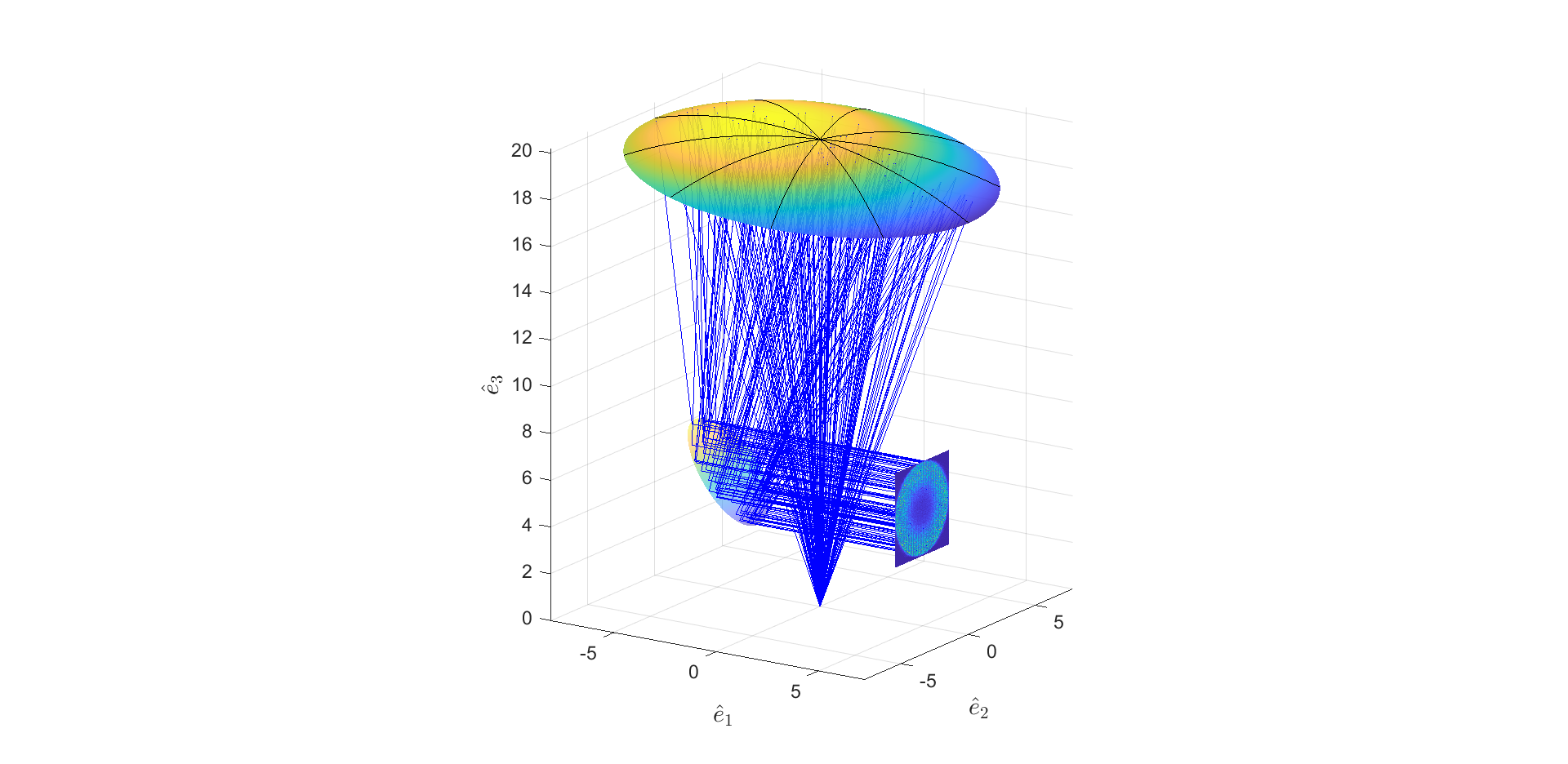}
  \caption{The optical system of our solution to the laser diode to ring-target problem.}\label{fig:donut sys}
\end{figure}

The second test case for our algorithm is the case of a laser diode \cite{sun15} to a ring-shaped target pattern. In this case we choose $\varphi=\pi/2$ and $\theta=0$, so that the outgoing beam is parallel to the $\bm{\hat{e}}_1$-axis. Note that the angle $\theta$ only matters when the source distribution is not rotationally symmetric. The intensity distribution of a laser diode can be modeled by an elliptical Gaussian on a plane perpendicular to the $\bm{\hat{e}}_3$-axis \cite{li92}. The emitted light has a $1/e^2$ intensity angle of $\theta_x$ in the $\bm{\hat{e}}_1,\bm{\hat{e}}_3$-plane, and $\theta_y$ in the $\bm{\hat{e}}_2,\bm{\hat{e}}_3$-plane. From this we can derive a density in terms of the $\bm{x}$-variables, see App. \ref{sec:diode}. The following density function is obtained:
\begin{equation}\label{eq:gauss source}
  f(\bm{x}) = B_x \frac{1+|\bm{x}|^2}{(1-|\bm{x}|^2)^3} \left(\exp\left[\frac{-8}{(1-|\bm{x}|^2)^2}\left(\frac{x_1^2}{\sigma_x^2}+\frac{x_2^2}{\sigma_y^2}\right)\right]+\delta\right),
\end{equation}
with $\sigma_x=\tan(\theta_x/2)$ and $\sigma_y=\tan(\theta_y/2)$. The scalar $B_x$ is a scaling parameter such that the flux of $f$ over $\Xc$ is 1. The parameter $\delta$ is used to ensure a minimal value for the source density. The source domain $\Xc$ is given by
\begin{equation}
  \Xc = \left\{\bm{x}\in\mathbb{R}^2 \mid \rho\leq\tan\left(\Theta/2\right)\right\},
\end{equation}
where $\Theta=\max(\theta_x,\theta_y)/2$. \\
\indent We want to have a target which consists of a ring on a target plane with an outer radius $r_o$ and an inner radius $r_i$. For this we will construct a density as a function of the unscaled target variables $\bm{y}$. We want this density to be uniform on a ring $\Omega_r$. With our algorithm we need a simply connected domain. So, instead of a ring, the domain we use will be a disk $\Omega_T=\Omega_r\cup\Omega_i$, where $\Omega_i$ is the circular domain enclosed by $\Omega_r$. The ring will have a higher density than the inner disk. Our model assumes that the density is smooth, so we will need to approximate this discontinuous change in density between the ring and the inner circle. From the derivation in App. \ref{sec:ring} we obtain the target intensity distribution
\begin{equation}
  g(\bm{y}) = \frac{c}{1+e^{-2k(|\bm{y}|-r_i)}}+\varepsilon.
\end{equation}
We choose $c$ so that the total flux over the target domain is equal to 1. The parameters $k$ and $\varepsilon$ determine how close this density is to the ideal density, uniform on $\Omega_r$ and zero on $\Omega_i$. They also influence the convergence of the algorithm. Generally a smaller $k$ or larger $\varepsilon$ will give better convergence, but a less pronounced difference in density between the inner region and the outer ring. \\
\indent For the source density, we choose to work with $1/e^2$ intensity angles $\theta_x=45^\circ$ and $\theta_y=13^\circ$. These are typical values for a laser diode \cite{sun15}. To avoid a too large difference in intensity in the source, we choose $\delta=10^{-3}$ in the density function from Eq. (\ref{eq:gauss source}). The target domain consists of an outer radius $r_o=2$ and an inner radius $r_i=1$. This is shifted by $-5$ along the $\bm{\hat{a}}_1$-axis in the target plane at a distance $l=5$ from the point source. We choose $\varepsilon$ such that the density in the inner circle is at least $10\%$ of the density in the ring. Furthermore, we used $k=100$ in the target density. We set $\beta=30$ and $u_0=20$. Because the intensity on the boundary of the source is much lower than at the center, it is more difficult for the algorithm to find a good mapping for the boundary compared to our previous test case. To counter this problem we decrease $\alpha$. We now choose $\alpha=0.01$ to put more emphasis on $J_B$ relative to $J_I$. Compared to the previous cases, we also increase the number of grid points and apply a $400\times400$ grid to the source domain. Tracing one million rays through the system that results from our algorithm gives us the illumination pattern and optical system in Fig. \ref{fig:donut illum} and  Fig. \ref{fig:donut sys}.

\section{Conclusion}
In this paper we introduced a method for computing the shapes of two reflectors to collimate a beam from a point source, for given source and target light distributions. A specific point of interest is the fact that the outgoing beam can be in any arbitrary direction. First we have derived a relation between the surface shapes and locations, and the optical mapping. Then we derived a Monge-Amp\`ere type equation for this problem. We proposed an algorithm based on the least-squares method to solve this equation. \\
\indent We tested our algorithm on two test cases. A uniform source and target, and a model of a laser diode to a ring-shaped target. The former was used to test the algorithm and the different layouts that it could attain. The latter consisted of more complicated source and target distributions instead. The second test case showed us that we might need to put some restrictions on our model to improve the convergence of the algorithm. For example, the variation of the intensity in the source cannot be too large. This was solved by imposing a minimal value of the source distribution. \\
\indent For further research there are several points of interest. In this paper we have skipped over some practical constraints. There is nothing in the algorithm yet preventing the rays from crossing a reflector. For example, the first reflector might (partially) be in the way of the outgoing beam. We managed to work around this by varying parameters to get a feasible solution. However, for the future it might be interesting to research if it could be possible to incorporate physical constraints like this into our algorithm. Furthermore we want to extend our algorithm to include more physical phenomena. Examples of this are Fresnel reflection (for lenses) and scattering.

\section*{Acknowledgements}
We thank Stefan B\"aumer for helping us with the example of a laser diode and a ring-shaped target.

\section*{Disclosures}
The authors declare no conflicts of interest.

\appendix
\section{Laser diode model}\label{sec:diode}
The intensity distribution of a laser diode can be modeled by an elliptical Gaussian on a plane perpendicular to the $\bm{\hat{e}}_3$-axis \cite{li92}. We choose the plane $z=z_p$. The intersection of a ray with this plane is given by the position vector $\bm{p}=(p_1,p_2)\tran$ on this plane. The $p_1$- and $p_2$-axes are parallel to the $\bm{\hat{e}}_1$- and $\bm{\hat{e}}_2$-axes, respectively. The intensity on the plane is then given by
\begin{equation}
  I(\bm{p}) = B \exp\left[-2\left(\frac{p_1^2}{w_1^2}+\frac{p_2^2}{w_2^2}\right)\right],
\end{equation}
where $w_1$ and $w_2$ are the $1/e^2$ intensity radii. These are the values along the $p_1$- and $p_2$-axes where the intensity has decreased to a factor $1/e^2$ of the maximum. The factor $B$ will be used to scale the function such that the total flux of the source is equal to 1. \\
\indent We denote by $\Omega_p$ the source domain in $\bm{p}$-coordinates, so $\Omega_p=\left\{\bm{p}=\bm{p}(\bm{x}) \mid \bm{x}\in\Xc\right\}$. We need to have a finite support for the intensity, so we choose the domain $\Omega_p$ given by $|\bm{p}|\leq\max\big(w_1,w_2\big)$. The difference between $w_1$ and $w_2$ can often be quite large. This could create a ratio of many orders of magnitude between the minimal and maximal value of $I$ in the domain. This causes problems with grid lines getting too close to each other on the target. Therefore we add a constant $\delta>0$ to the intensity to assure a minimal value. This gives the new intensity
\begin{equation}\label{eq:tildeI}
  \tilde{I}(\bm{p}) = \tilde{B}\left( \exp\left[-2\left(\frac{p_1^2}{w_1^2}+\frac{p_2^2}{w_2^2}\right)\right] + \delta \right).
\end{equation}
Again, the factor $\tilde{B}$ is used to scale the total flux to 1. Now we have a formula for the source intensity on the plane $z=z_p$, but we want to convert this into an intensity $f$ in stereographic coordinates. For this, we first convert $\bm{p}$ to a directional vector $\sray$ and then to stereographic coordinates $\bm{x}$ to obtain
\begin{equation}\label{eq:p to x}
  \bm{p} = \frac{z_p}{s_3}\begin{pmatrix} s_1 \\ s_2 \end{pmatrix} = \frac{2z_p\bm{x}}{1-|\bm{x}|^2}.
\end{equation}
We already used $\Xc$ for the source domain in stereographic coordinates. For any subset $\mathcal{A}\subset\Xc$ we have a corresponding set $\mathcal{A}_P=\bm{p}\big(\mathcal{A}\big)\subset\Omega_p$. The total flux in those two sets should be equal, so
\begin{equation}
  \iint_{\mathcal{A}} f(\bm{x})\,\diff A(\bm{x}) = \iint_{\mathcal{A}_p} \tilde{I}(\bm{p})\,\diff A(\bm{p}).
\end{equation}
We write the right-hand side as an integral over $\mathcal{A}$ by using integration by substitution. Because this holds for any subset $\mathcal{A}$, this gives a source density function $f$, with
\begin{equation}
\begin{aligned}
  f(\bm{x}) &= \det\big(\Diff \bm{p}(\bm{x})\big)\tilde{I}\big(\bm{p}(\bm{x})\big) \\
  &= 4\tilde{B}z_p^2\frac{1+|\bm{x}|^2}{(1-|\bm{x}|^2)^3} \left(\exp\left[\frac{-8z_p^2}{(1-|\bm{x}|^2)^2}\left(\frac{x_1^2}{w_1^2}+\frac{x_2^2}{w_1^2}\right)\right]+\delta\right).
\end{aligned}
\end{equation}
This density seems to be dependent on $z_p$, while the intensity of the point source should of course not depend on the plane of projection. We will show that we can in fact write $f$ as a function independent of $z_p$. In general, the increase of $w_1$ and $w_2$ for increasing $z_p$ is given by the full angles at the source point, $\theta_x$ and $\theta_y$. So these radii of a laser diode increase linearly in $z_p$. We then have
\begin{equation}
  w_1 = z_p\tan(\theta_x/2),\qquad w_2 = z_p\tan(\theta_y/2),
\end{equation}
so that
\begin{equation}\label{eq:gauss source2}
  f(\bm{x}) = B_x \frac{1+|\bm{x}|^2}{(1-|\bm{x}|^2)^3} \left(\exp\left[\frac{-8}{(1-|\bm{x}|^2)^2}\left(\frac{x_1^2}{\sigma_x^2}+\frac{x_2^2}{\sigma_y^2}\right)\right]+\delta\right),
\end{equation}
with $\sigma_x=\tan(\theta_x/2)$, $\sigma_y=\tan(\theta_y/2)$ and $B_x=4\tilde{B}z_p^2$. With Eq. (\ref{eq:tildeI}) and integration by substitution, $B_x$ can be written as
\begin{equation}
  B_x = \left(\iint_{\Xc} \frac{1+|\bm{x}|^2}{(1-|\bm{x}|^2)^3} \left(\exp\left[\frac{-8}{(1-|\bm{x}|^2)^2}\left(\frac{x_1^2}{\sigma_x^2}+\frac{x_2^2}{\sigma_y^2}\right)\right]+\delta\right)\,\diff A\right)^{-1}.
\end{equation}
This shows that $B_x$ is independent of $z_p$ and indeed also our source density $f$ is independent of $z_p$. The only part left for our source density is to define the boundary of the domain in stereographic coordinates. We mentioned before that the boundary of the domain in $\bm{p}$-coordinates on the plane $z=z_p$ is given by $|\bm{p}|=\max\big(w_1,w_2\big)$, which is written as $|\bm{p}|=\max\big(z_p\tan(\theta_x/2),z_p\tan(\theta_y/2)\big)=z_p\tan\left(\Theta/2\right)$, with $\Theta=\max\big(\theta_x,\theta_y\big)/2$. From Eq. (\ref{eq:p to x}) we obtain
\begin{equation}
  |\bm{x}|=\frac{\sqrt{z_p^2+|\bm{p}|^2}-z_p}{|\bm{p}|}.
\end{equation}
On the boundary this gives
\begin{equation}
  |\bm{x}| = \frac{\sqrt{z_p^2+z_p^2\tan^2\left(\Theta\right)}-z_p}{z_p\tan\left(\Theta\right)} = \frac{1-\cos\left(\Theta\right)}{\sin\left(\Theta\right)} = \tan\left(\Theta/2\right).
\end{equation}
Again, this result is independent of the choice of $z_p$. We have now modeled the laser diode by the intensity distribution in Eq. (\ref{eq:gauss source2}), defined on the domain
\begin{equation}
  \Xc = \left\{\bm{x}\in\mathbb{R}^2 \mid |\bm{x}|\leq\tan\left(\Theta/2\right)\right\}.
\end{equation}

\section{Ring-target density}\label{sec:ring}
We define our circular domain $\Omega_T$ by an angle $\xi\in[0,2\pi)$ and a radius $r\in[0,r_o]$. The outer ring is denoted by $\Omega_r$ and given by $r\in[r_i,r_o]$. The inner disk $\Omega_i$ is then given by $r\in[0,r_i]$ such that we have $\Omega_T=\Omega_r\cup\Omega_i$. We denote the area of the ring by $A$ and we have
\begin{equation}
  A = \pi(r_o^2-r_i^2).
\end{equation}
Ideally, we would have a density function $\tilde{h}$ defined on $\Omega_T$ with values
\begin{equation}\label{eq:ideal}
  \tilde{h}(\bm{y}) = \begin{cases}
                1/A, & \mbox{for } |\bm{y}|\geq r_i,\\
                0, & \mbox{for } |\bm{y}|<r_i.
              \end{cases}
\end{equation}
This density has similarities to the Heaviside stepfunction $H$. We use this to write $\tilde{h}(\bm{y})=(1/A)H(|\bm{y}|-r_i)$. As can be seen in Eq. (\ref{eq:Monge polar-a}), we cannot have a target intensity equal to 0 at any point in the target domain. To avoid this, we construct a density which has value $\varepsilon$ on $\Omega_i$. We define the function $h$ by
\begin{equation}\label{eq:ideal density}
  h(\bm{y}) = cH(|\bm{y}|-r_i)+\varepsilon = \begin{cases}
                                          c+\varepsilon, & \mbox{for } |\bm{y}|\geq r_i, \\
                                          \varepsilon, & \mbox{for } |\bm{y}|<r_i.
                                        \end{cases}
\end{equation}
The problem with the density function that we have proposed now is that it is discontinuous across the circle $|\bm{y}|=r_i$. Experiments have shown that our algorithms will not work with discontinuities. To avoid this problem, we have to approximate the discontinuous density by a smooth function. The Heaviside stepfunction that was mentioned before can be approximated by
\begin{equation}
  H(y) \approx \frac{1}{2}\big(1+\tanh(ky)\big) = \frac{1}{1+e^{-2ky}},
\end{equation}
where $k>0$. The approximation converges pointwise to $H$ for $k\to\infty$ \cite[Ch.~9]{old10}. We use this in combination with Eq. (\ref{eq:ideal density}) to construct a density that approximates the ring target. This density is given by
\begin{equation}
  g(\bm{y}) = \frac{c}{1+e^{-2k(|\bm{y}|-r_i)}}+\varepsilon.
\end{equation}

\bibliographystyle{unsrt}
\bibliography{bib1}

\end{document}